\documentclass[twocolumn,showpacs,preprintnumbers,amsmath,amssymb]{revtex4}
\usepackage{graphicx}% Include figure files
\usepackage{amssymb}
\usepackage{dcolumn}% Align table columns on decimal point
\usepackage{bm}% bold math
\usepackage{epstopdf}

\begin{document}

\preprint{APS/123-QED}

\title{  States near Dirac points  of  rectangular graphene dot in a magnetic field}
\author{S. C. Kim$^{1}$, P. S. Park$^{1}$, and S.-R. Eric Yang$^{1,2}$\footnote{ corresponding author, eyang@venus.korea.ac.kr}}
\affiliation{$^{1}$Physics Department, Korea  University, Seoul Korea } \affiliation{$^{2}$Korea
Institute for Advanced Study, Seoul Korea }
\date{\today}

\begin{abstract}
In neutral graphene dots the Fermi level coincides with the Dirac points. We have investigated in
the presence of a magnetic field several unusual properties of single electron states  near the
Fermi level of such a rectangular-shaped graphene dot with two zigzag and two armchair edges. We
find that a quasi-degenerate level forms near zero energy and the number of states in this level
can be tuned by the magnetic field. The wavefunctions of states in this level are all peaked on the
zigzag edges with or without some weight inside the dot. Some of these states are  magnetic
 field-independent
surface states while the others are field-dependent. We have  found
a scaling result from which  the number of  magnetic field-dependent
states of large dots can be inferred from those of smaller dots.

\end{abstract}


\maketitle

\section{Introduction}

\begin{figure}[!hbpt]
\begin{center}
\includegraphics[width=0.25 \textwidth]{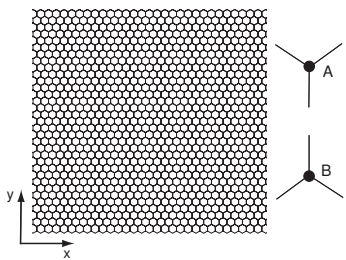}
\caption{Finite graphene layer with  zigzag and armchair edges. There are equal number of A and B
carbon atoms.
The graphene layer has reflection symmetries about horizontal and vertical lines that go through
the center of the layer.
A magnetic field is present perpendicular to the layer.
} \label{fig:20site}
\end{center}
\end{figure}

Graphene dots have a great potential for many applications since they are the elemental blocks to
construct graphene-based nano devices. It is possible to cut graphene sheet\cite{Geim} in the
desired shape and size\cite{Sci}, and use it to make quantum dot  devices. In such devices it
may be  possible to realize  experimentally zigzag or armchair boundaries.

Graphene systems possess several unusual physical properties associated with the presence of the
Dirac points. For example, compared to ordinary Landau levels of quasi-two-dimensional
semiconductors the lowest Landau level (LLL) of graphene  is peculiar since it has  {\it zero}
energy that is independent of magnetic field\cite{Mc,Zhang,Novo,Zh,Sh}. Moreover, wavefunctions
of the LLL are {\it chiral}, i.e., the probability amplitude of find the electron on one type
carbon atoms is zero. There are other graphene systems  with  zero energy states.
Semi-infinite\cite{Fu} or nanoribbon graphene\cite{Brey, Son} with zigzag edges along the x-axis
develop a  flat band of {\it zero} energy chiral states. These states are surface states and are
localized states at the zigzag edges with various localization lengths\cite{Fu}. The zigzag edge
and  the LLL states have zero energy because their wavefunctions are chiral. Effects of a magnetic
field on graphene Hall bars have been investigated recently, and some zero energy chiral states are
found to be strongly localized on the zigzag edges in addition to the usual LLL states\cite{Brey2,Cas,Gu,Ari}.

One may expect that the degeneracy of chiral  states with zero energy will be split when quantum
confinement effect is introduced in a  graphene dot\cite{Marti,Gia,Haus}. However, the splitting of
these energies may be unusual
in some graphene dots\cite{Sch,ZZZ,Rec,Gutt}. Recently the magnetic field dependence of these levels
in a gated graphene dot was investigated experimentally\cite{Gutt}.
Effects of various types of edges have been  also investigated: zigzag-edged dots,
armchair-edged dots\cite{Fer,Ez,Od,ZZZ},
and rectangular graphene dots with two zigzag  and two armchair edges\cite{Bj,Cha}
have been studied.
Armchair edges
couple states near K and K$^\prime$ points of the first  Brillouin zone and
 generate
several mixed chiral zigzag edge states with {\it nearly}  zero energies.  In the
rectangular dots the number of these states, $ N_l$,  may be determined from the condition that the
x-component of wavevectors satisfies $1/L_y<k_{x,n}<\pi/3a$, where $k_{x,n}=\frac{\pi
n}{L_x}-\frac{2\pi}{3a}$, $a=\sqrt{3}a_0$ is the length of the unit cell, and $n=0,\pm1,\pm2,...$
(the  nearest neighbor carbon-carbon distance is
$a_0=1.42\AA$, the  horizontal length of the dot is $L_x=\sqrt{3}Ma_0$ with  $M$  number of hexagons along the x-axis, and
the vertical length is  $L_y=a_0(3N+2)$  with
$N$  the number of hexagons and  $(N+1)$ carbon  bonds along the  y-axis. See Fig.\ref{fig:20site}).
 This condition  implies that the integer $n$ is given by
\begin{eqnarray}
\frac{\sqrt{3}M}{(3N+2)\pi}+\frac{2M}{3}\leq n \leq M.
\label{Eq:num}
\end{eqnarray}
The effective mass approximation wavefunctions of these surface states are derived  in Ref.\cite{Cha}.

\begin{figure}
\begin{center}
\includegraphics[width=0.25\textwidth]{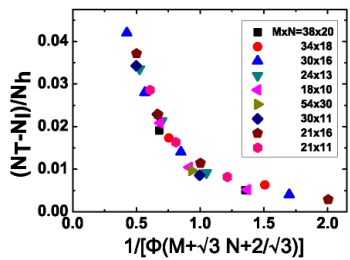}
\caption{Number of nearly zero energy  states that are induced by a
magnetic field follow a scaling curve when plotted as a function of
$\frac{1}{(M+\sqrt{3}N+2/\sqrt{3})\phi}$ for different rectangular
shaped graphene dots.  Here $\delta=0.01 eV$.
} \label{fig:scaling}
\end{center}
\end{figure}

We investigate how  properties of rectangular dots change in the
presence of a magnetic field in the regime where
Hofstadter-butterfly effects\cite{Sch,ZZZ,Gutt} are negligible. In
neutral graphene dots the Fermi level has zero energy, and,
consequently, magnetic, optical, and STM properties are expected to
depend  on the number of available states near zero energy. Our
investigation shows that  the wavefunctions of nearly zero energy
states are all  peaked on  the zigzag edges with or without
appreciable weight inside the dot. This may be understood as  mixing
of LLL and surface states by armchair edges  of the square dot
through  intervalley scattering, which is unique to the square dot
(This will be explained in Sec.IV).

Our
study shows that the number of  states within the energy interval $\delta$ around zero energy is
given by
\begin{eqnarray}
N_T(\phi)= N_l+N_D(\phi)
\label{eq:NT}
\end{eqnarray}
(the energy $\delta$ is typically less than  the quantization energy of a rectangular dot,
which can be estimated using the Dirac equation:
$\gamma k_{min}\sim t\frac{a}{L_{x,y}}$, where $\gamma=\sqrt{3}ta/2$ with the  hopping energy $t$).
Here magnetic edge states are not included since their energies are larger than $\delta$.
$N_D(\phi)$ is the number states at zero magnetic field that merge into  the energy interval
$\delta$ around zero energy as the dimensionless magnetic flux $\phi=\frac{\Phi}{\Phi_0}$
increases.
This effect provides a means to control
the number of states at the Fermi level.
There are other  states with nearly zero energies at
$\phi=0$, which  remain so even at $\phi\ne 0$.
The localization lengths of these states are shorter than the magnetic length. We denote the
number of these states  by
$ N_l$.

Our numerical results indicate that for relatively small rectangular-shaped graphene dots with size less than of order $10^2\AA$
the number of states at the Fermi level display a negligible magnetic field dependence for
values
that are usually accessible experimentally
($B<10T$ corresponds, in dimensionless magnetic
flux, to $\phi=\frac{\Phi}{\Phi_0}< 10^{-4}$, where
$\Phi_0=hc/e$ and   $\Phi=BA_{h}$ with the area of a hexagon $A_h=\frac{3\sqrt{3}}{2}a_0^2$).
%In our tight binding calculation the effect of magnetic field enters through the Peierls phase factor
%$t e^{i\frac{e}{\hbar}\int_{\vec{R}_j}^{\vec{R}_i}\vec{A}\cdot d\vec{r}}$.  The maximum value of the phase
%$\frac{e}{\hbar}\int_{\vec{R}_j}^{\vec{R}_i}\vec{A}\cdot d\vec{r}$
%is $\pi\phi m$, where
%the integer $m$ is of order $L_{x,y}/a$. For small dots the product $\pi \phi m$ is rather small and the effect of magnetic field is
%barely visible.  This argument gives an estimate of an energy scale $t \pi \phi m$
On the other hand, in larger dots  this dependence is   significant.
However, it is computationally difficult to investigate large
rectangular-shaped graphene dots since the number of carbon  atoms increases rapidly with the size.
We have found a
scaling result from which  one can infer results for larger dots from those of smaller dots.
For different rectangular-shaped
graphene dots and values of $\phi$ it can be described well  by the following dimensionless form
\begin{eqnarray}
N_D(\phi)=\frac{L_xL_y}{A_h}f(\frac{\ell^2}{a_0(L_x+L_y)}),
\label{eq:scaling}
\end{eqnarray}
where $\ell=\frac{3^{3/4}a_0}{(4\pi\phi)^{1/2}}$ is the magnetic length and $f(x)$ is a
scaling function, see Fig.\ref{fig:scaling}. The total number of
hexagons in the dot is  $N_h=L_xL_y/A_h$.
%This scaling function may be used to infer the values of $N_D(\phi)$ for large rectangular dots that
%require heavy numerical computing.
Our numerical result
shows that the dependence of $N_D(\phi)$ on $\phi$ is initially   non-linear
in the regime where  the diameter of the cyclotron motion is comparable
to the system length,  $2 \ell \sim L_{x,y}$.

\section{ Number of  states in the quasi-degenerate level}

Our Hamiltonian is
\begin{eqnarray}
H&=&-\sum_{<i,j>}t_{ij}c^{\dagger}_{i}c_{j},
\end{eqnarray}
where $t_{ij}=te^{i\frac{e}{\hbar}\int_{\vec{R}_j}^{\vec{R}_i}\vec{A}\cdot d\vec{r}}$ are the
hopping parameters and $c^{+}_{i}$ creates an electron at site $i$.
Here we use a Landau gauge $\vec{A}=B(-y,0,0)$. The summation $<i,j>$ is over nearest neighbor
sites and $t=2.5$eV. The eigenstate with eigenenergy $\epsilon_n$ is denoted by
$\phi_{\epsilon_n}(\vec{R})$, where $\vec{R}$ labels each lattice point. Because of electron-hole
symmetry eigenvalues appear in pairs of positive and negative values $(\epsilon,-\epsilon)$, and
the probability wavefunctions of a pair of states, $(|\phi_{\epsilon_n}(\vec{R})|^2,
|\phi_{-\epsilon_n}(\vec{R})|^2)$, are identical. Our numerical results are consistent with this.
%In this work we choose $L_x=73.79\AA$ and $L_y=71\AA$ ($M=30$ and $N=16$) unless stated otherwise
%(in the investigation of the scaling results we have used larger dots with  $L_{x,y}$ up to $130\AA$).

\begin{figure}
\begin{center}
\includegraphics[width=0.25\textwidth]{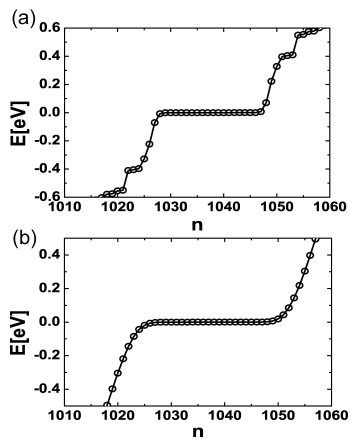}
\caption{(a) Eigenenergies $\epsilon_n$ at  $\phi=0$.
Quasi-degenerate states are present near zero
energy.
Size of the dot is $74\times 71\AA^2$.
A quantization energy of order $\gamma k_{min}\sim 0.03$eV can be seen as an excitation gap  near zero energy.
(b) Eigenenergies $\epsilon_n$ at  $\phi=0.01$.
} \label{fig:energyzeroflux}
\end{center}
\end{figure}

\begin{figure}
\begin{center}
\includegraphics[width=0.25\textwidth]{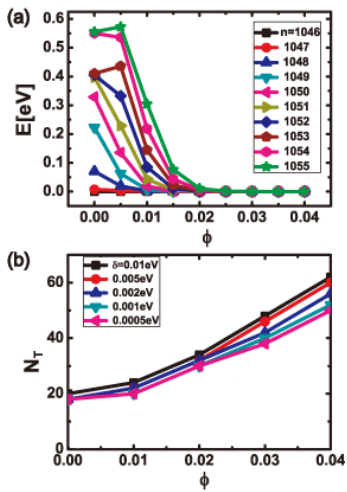}
\caption{(a) Some energy levels $\epsilon_n$
change  as $\phi$ increases  while some do not. (b) The total number of
states within the energy interval $\delta$ around zero energy at a
finite value of $\phi$. Size of the dot is $74\times 71\AA^2$.
}\label{fig:energychangeflux}
\end{center}
\end{figure}

\begin{figure}
\begin{center}
\includegraphics[width=0.35\textwidth]{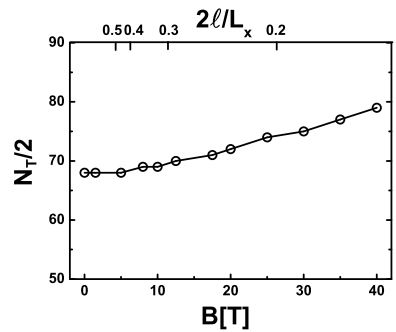}
\caption{Results for a dot with size  $50\times50 nm^2$.
%(a) The probability wavefunction of the state with $\epsilon_{3362}=-4.5\times10^{-3}$eV at  $B=0T$.
%(b) The probability wavefunction of the same state  at  $B=200T$ ($\epsilon_{3362}=-1.1\times10^{-3}$eV).
%The ratio is $\ell/L_x=0.15$.
Dependence of $N_T/2$ on $B$ for $\delta=0.01$eV.
}\label{fig:large}
\end{center}
\end{figure}

Figs.\ref{fig:energyzeroflux}(a) and (b) display the energy spectra
near zero energy  for $\phi=0$ and $0.01$. At  $\phi=0$ there are
approximately $20$ states within $|\epsilon_{n}|<0.01eV$, consistent
with the analytical result of Eq.(\ref{Eq:num}). At $\phi=0.01$ the
numerical  value is increased to $24$.
Fig.\ref{fig:energychangeflux}(a) shows how some energy levels
$\epsilon_n$ at $\phi=0$ change as a function of the magnetic flux
$\phi$. These energy levels do not anticross. We observe that nearly
zero energies at $\phi=0$ do not change noticeably in magnitude as
$\phi$ varies. There  are $ N_l$ such localized surface states. On
the other hand, we find that as $\phi$ increases non-zero energies
become smaller and move closer to zero. This implies that, for a
given energy interval $\delta$, the number of states in it,
$N_T(\phi)$, increases with $\phi$. From
Fig.\ref{fig:energychangeflux}(b) we see that it displays a
non-linear dependence on $\phi$. For a large dot of size
$50\times50nm^2$ a similar  dependence of  $N_T$ on $B$ is seen, as
shown in Fig.\ref{fig:large}.  Non-linear dependence occurs in the
regime $2\ell/L_x \sim  0.5$. As a test of our numerical procedures
we have verified that the sum of  $N_D(\phi)=N_T(\phi)- N_l$ and the
number of  magnetic edge states is equal to the total bulk Landau
level degeneracy $2D_B$
($D_B=\frac{2L_{x}L_{y}}{3\sqrt{3}a_{0}^2}\phi$ is the degeneracy
per valley).

The ratio between the number of nearly zero energy states induced by the magnetic field and the number of hexagons,
$\frac{N_D(\phi)}{N_h}$, should depend on a dimensionless quantity
consisting of a combination of $\ell$, $a_0$, $L_x$ and $L_y$, which are the important parameters of rectangular
graphene dots.
%The carbon-carbon distance  $a_0$ appears in the scaling variable  because it represents  the minimum localization length
%of surface states\cite{Cas}.
The lengths
$L_x$ and $L_y$
should appear as  $L_x+L_y$ so that  for rectangular-shaped graphene sheets $N_D/N_h$ remains the
same when $L_x$ and $L_y$ are exchanged.
These considerations lead us to  the dimensionless variable  $\frac{\ell^2}{a_0(L_x+L_y)}=\frac{3}{4\pi}
\frac{1}{\phi (M+\sqrt{3}N+2/\sqrt{3})}$, see Eq.(\ref{eq:scaling}).
We are especially interested  in the regime where the diameter of the cyclotron orbit is comparable to
the system length
$2\ell\sim L_{x,y}$.
Note that in this regime
many cyclotron orbits
get affected by  the presence of the edges and corners of the rectangular dot.
%In the opposite limit  $2\ell/ L_{x,y}\rightarrow 0$, where only a vanishingly small fraction of these orbits
%experiences the edges and corners, a different scaling function of the bulk LL degeneracy with
%$N_D(\phi)\propto \phi$ should be used.
Since  we must also assume that  Hofstadter effect is negligible the
validity regime of Eq.(\ref{eq:scaling}) is $a_0\ll 2\ell<L_{x,y}$.
%(the available  numerical data that are used to find the scaling function are
%in this regime).
Note also that the scaling function $f(x)$ should be different for each $\delta$.

\section{ Wavefunctions of quasi-degenerate states in a magnetic field }

\begin{figure}
\begin{center}
\includegraphics[width=0.2\textwidth]{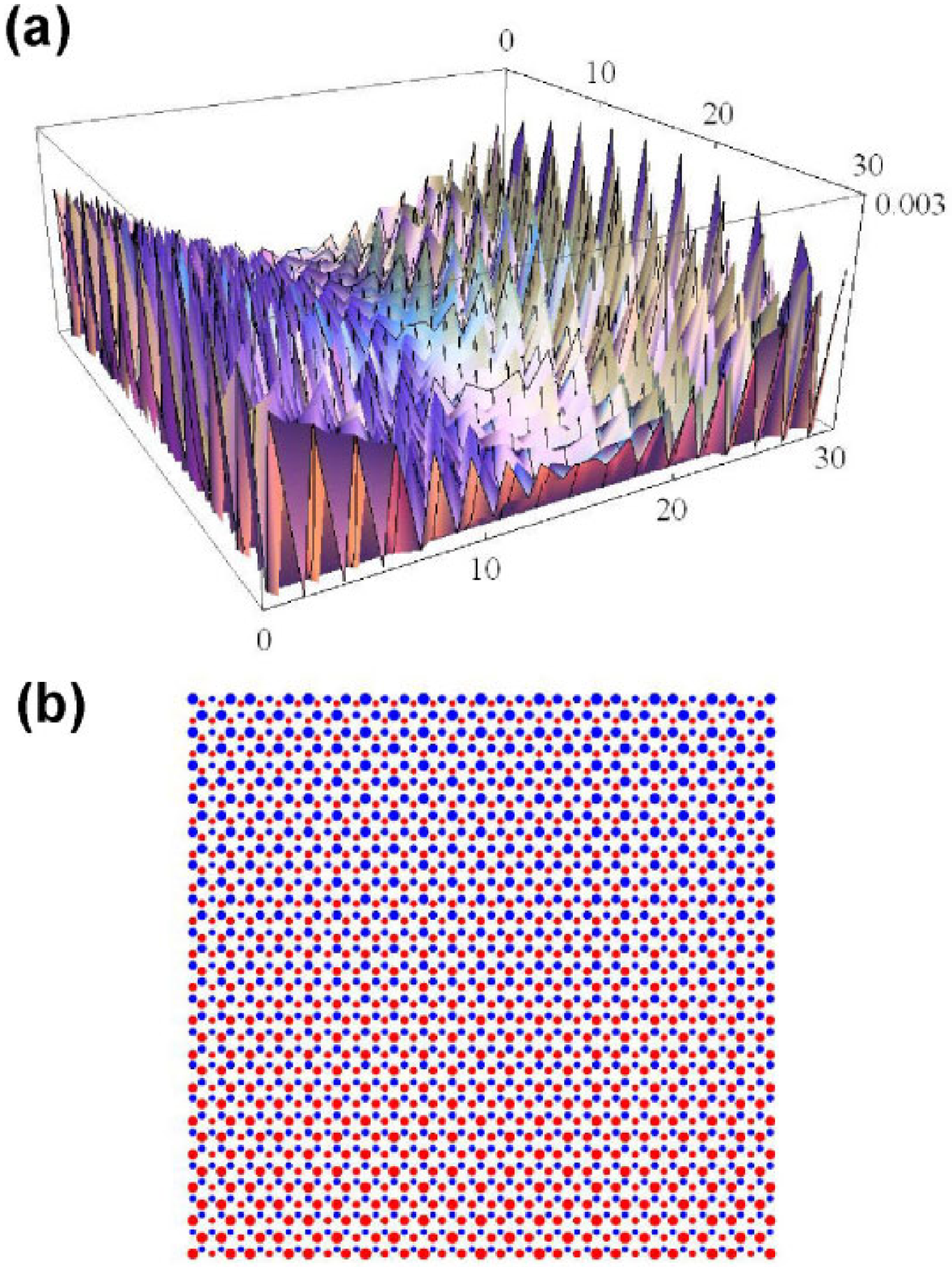}
\includegraphics[width=0.2\textwidth]{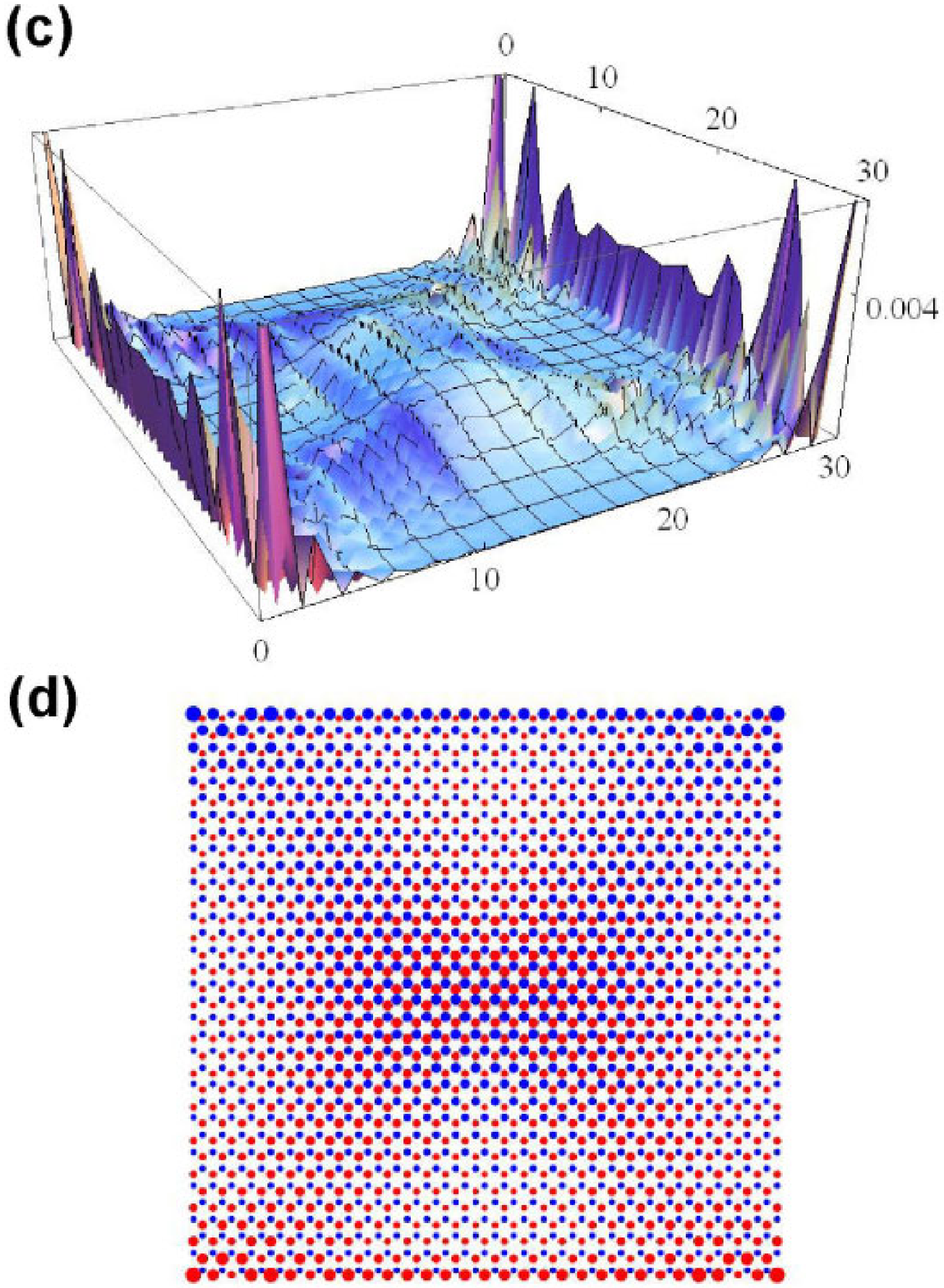}
\caption{
Size of the dot is $74\times 71\AA^2$.
(a) The probability wavefunction of the state with
$\epsilon_{1027}=-0.07$eV at $\phi$=0. The length unit is $a$. (b)
Profile of z-component of pseudospin: sizes of red  (blue) dots represent probabilities of occupying A
(B) carbon atoms.  Note that blue  (red) dots are dominant in the upper (lower) of
 dot.  When the probabilities  are less than 0.00001, the
radius of the dots is set to the smallest value.
The upper and lower horizontal edges represent zigzag edges.
(c) The probability wavefunction of the state with $\epsilon_{1027}=-2.0\times 10^{-3}$eV at $\ell/L_x=0.12$ ($\phi=0.01$).
(d) Profile of z-component of pseudospin for $n=1027$ at  $\phi=0.01$.}
\label{fig:wavefunction1027zero}
\end{center}
\end{figure}

\begin{figure}
\begin{center}
\includegraphics[width=0.2\textwidth]{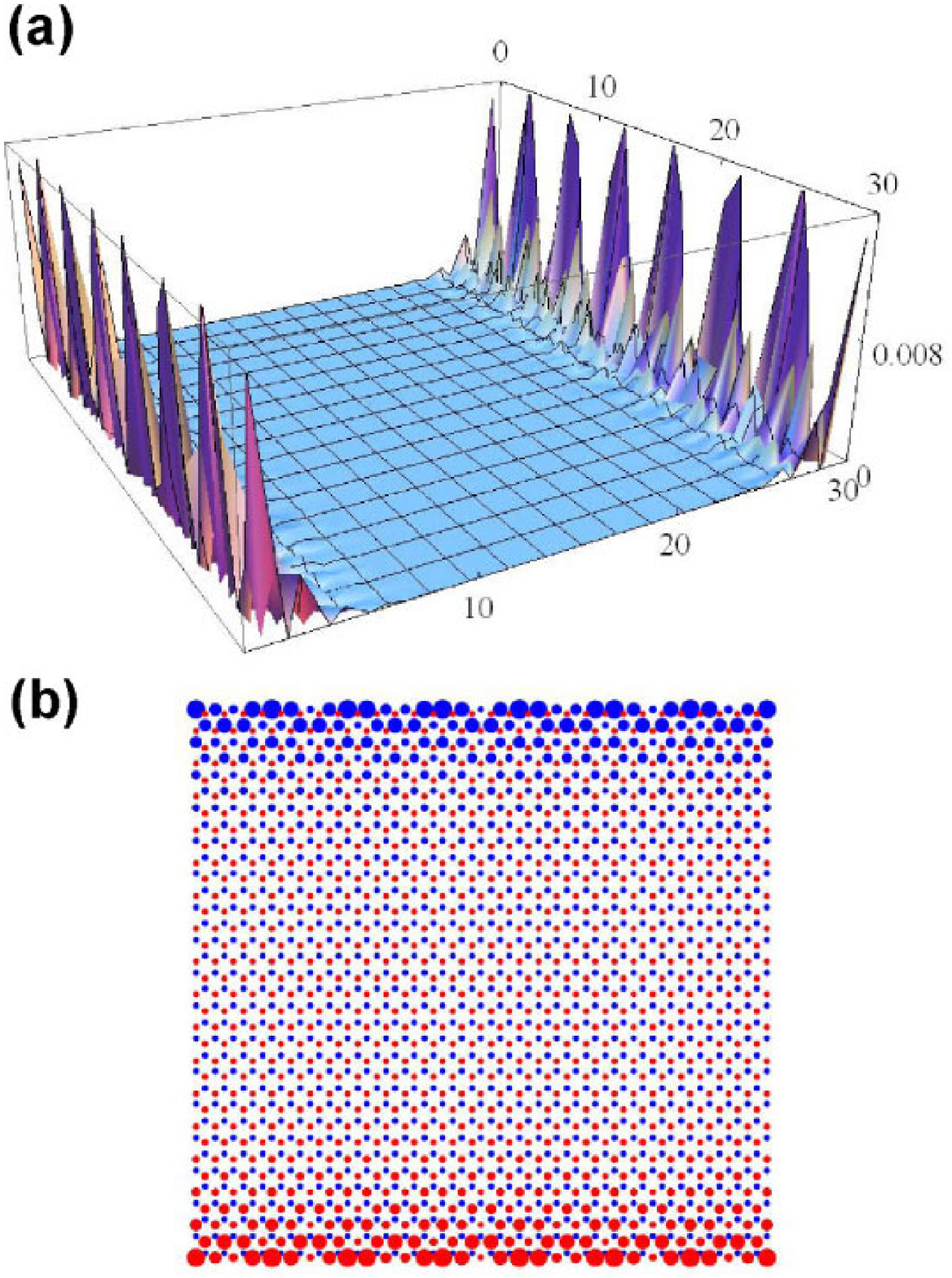}
\includegraphics[width=0.2\textwidth]{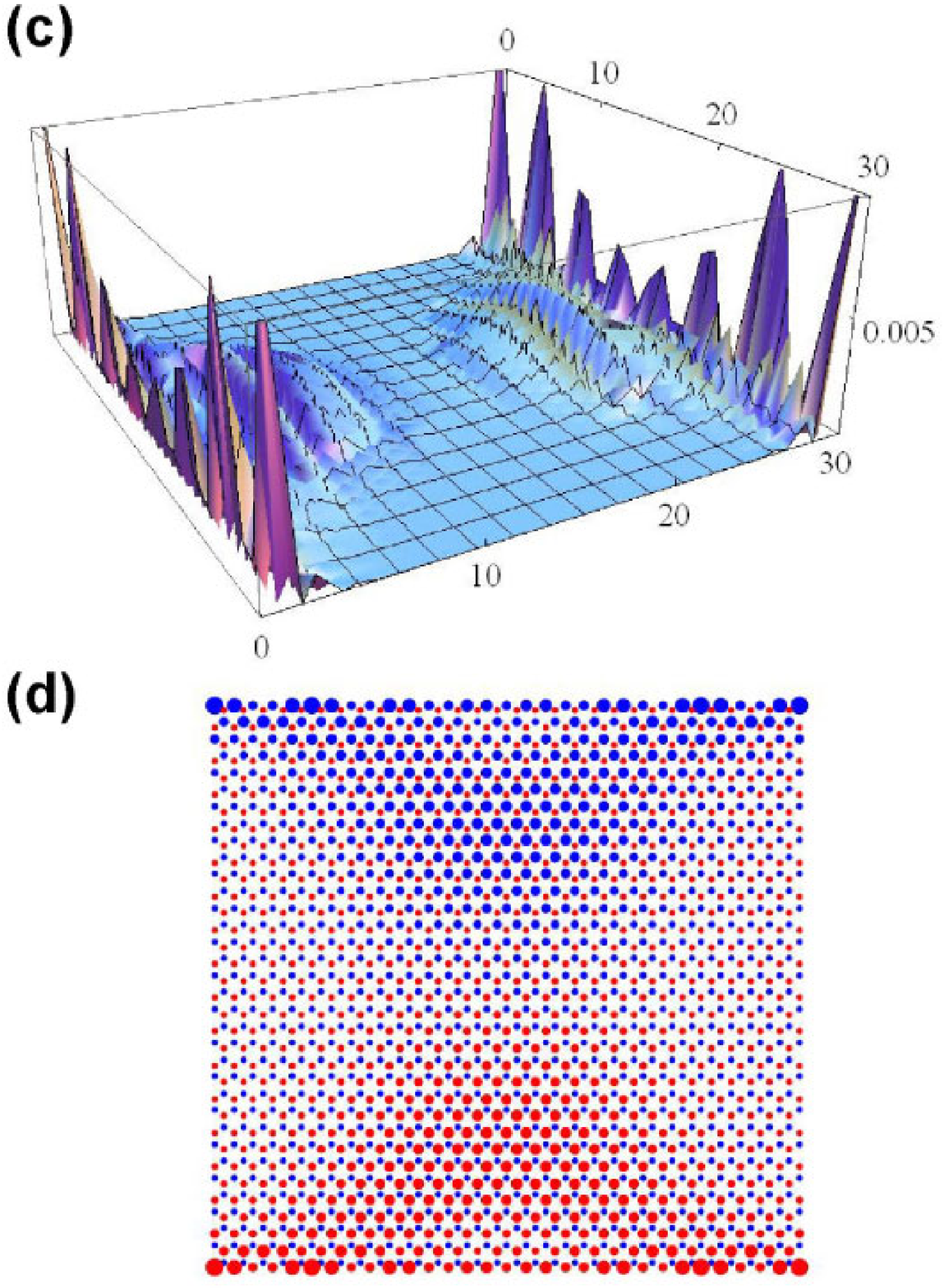}
\caption{
Size of the dot is $74\times 71\AA^2$.
(a) The probability wavefunction of the state with $\epsilon_{1045}=5.38\times10^{-6}$eV at
$\phi=0$.
(b) Profile of z-component of pseudospin for  $n=1045$ at $\phi=0$.
(c) The probability wavefunction for  $n=1045$ at $\ell/L_x=0.12$ ($\phi=0.01$).
(d) Profile of z-component of pseudospin for  $n=1045$ at $\phi=0.01$.}\label{fig:wavefunction1045zero}
\end{center}
\end{figure}

\begin{figure}
\begin{center}
\includegraphics[width=0.2\textwidth]{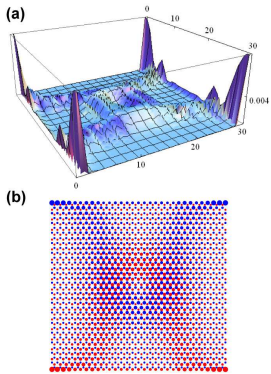}
\caption{Size of the dot is $74\times 71\AA^2$ and $\ell/L_x=0.09$ ($\phi=0.02$)
(a) The probability wavefunction of the state with $\epsilon_{1027}=-2.04\times10^{-5}$eV.
(b) Profile of z-component of pseudospin of the same state. These results should compared with those in
Fig.\ref{fig:wavefunction1027zero}.
}\label{fig:edge2}
\end{center}
\end{figure}

\begin{figure}
\begin{center}
\includegraphics[width=0.2\textwidth]{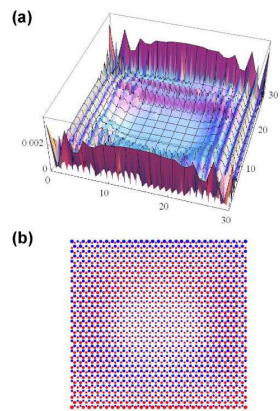}
\caption{(a) The probability wavefunction of the state with $\epsilon_{1053}=0.144$eV at
$\ell/L_x=0.12$ ($\phi=0.01$). Size of the dot is $74\times 71\AA^2$.
(b) Profile of z-component of pseudospin of the same state.}\label{fig:edge}
\end{center}
\end{figure}

We first show how the wavefunction of a non-zero energy state at
$\phi=0$ changes into a state with nearly zero energy as $\phi$
increases. Consider the probability wavefunction for $n=1027$ at a
finite $\phi=0.01$, as shown in
Fig.\ref{fig:wavefunction1027zero}(c). It is localized  on the
zigzag edges with a finite probability inside the dot. On the
armchair edges the  wavefunction is  vanishingly small. The
wavefunction has changed significantly from the $\phi=0$ result, see
Fig.\ref{fig:wavefunction1027zero}(a), and also its energy  has
changed from -0.07eV to $-2.0\times 10^{-3}$eV. The values of the
z-component of the pseudospin, Fig.\ref{fig:wavefunction1027zero}(b)
and  Fig.\ref{fig:wavefunction1027zero}(d), are larger on the zigzag
edges at $\phi=0.01$ compared to the result  at $\phi=0$. The
probability wavefunction of another state with nearly zero energy is
shown in Fig.\ref{fig:wavefunction1045zero}(c)  at  $\phi=0.01$. We
see that the result is somewhat different from the zero field result
of Fig.\ref{fig:wavefunction1045zero}(a), which displays a localized
state with the localization length comparable to the unit cell
length $a$. Now there is a finite probability to find an electron
inside the dot while the probabilities  on the zigzag edges are
reduced. Note that the energy of this state has changed from
$5.38\times 10^{-6}$eV to $4.5\times 10^{-6}$eV when $\phi$ is
changed to 0.01 from zero. The pseudospin profiles are shown in
Figs.\ref{fig:wavefunction1045zero}(b) and (d). Fig.\ref{fig:edge2}
 shows a probability wavefunction
at a smaller value of  $\ell/L_{x,y}=0.09$ (corresponding to $\phi=0.02$), and we see
that the wavefunction is less localized on the zigzag edges and LLL character
is more pronounced in comparison to  result of $\ell/L_x=0.12$ (Fig.\ref{fig:wavefunction1027zero}(c)).
All $N_D(\phi)$  states have similar properties mentioned above with   finite probabilities of finding an electron
inside the dot.
There are also  $N_{\ell}$ zigzag edges states that are more
strongly localized on the edges with  localization lengths comparable to $a$.
When an electron is in one of these states the probability of find the  electron away
from the edges is practically zero. The energies of these states are less then $10^{-10}$eV.
We can summarize our results as follows: all nearly
zero energy states are localized on  the zigzag edges  with
or without some   weight inside the dot.

%So far we have discussed effects seen in a smaller dot of size
% $74\times 71\AA^2$ at magnetic field values near  $B=1000$T.
%Similar effects can be seen in a larger dot with the size $13\times 13 nm^2$
%at a significantly lower
%magnetic field of order $100$T, see Fig.\ref{fig:large}.
%At this value of magnetic field the magnetic length is about $\ell=29\AA$ and  the ratio is $\ell/L_{x,y}=0.22$.
%The results of $N_D$ for this dot
%also follow the scaling curve of Fig.\ref{fig:scaling}.
%We estimate from scaling results that a more significant variation in $N_D$ may be observable at 1T for
%dots of size of order $100\times 100 nm^2$.

When the magnetic length is much smaller than the system size  magnetic edge states can be formed,
see Fig.\ref{fig:edge}. The probability wavefunction of a magnetic edge state with
$\epsilon_{1053}=0.144$eV is shown in Fig.\ref{fig:edge}(a). It is a mixture of ordinary magnetic
and zigzag edge states. The probability wavefunction decays from the armchair edges
while  it is strongly peaked on the zigzag edges. Pseudospin expectation values on
each  zigzag edge display  opposite chiral behavior. On armchair edges the chiralities  are more or
less evenly mixed, see  Fig.\ref{fig:edge}(b).

\section{ Discussions and conclusions}

\begin{figure}[!hbpt]
\begin{center}
\includegraphics[width=0.25\textwidth]{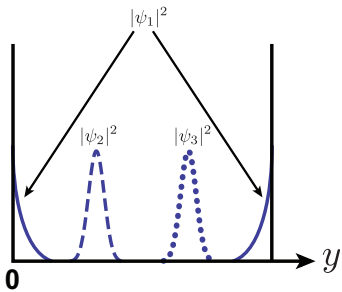}
\caption{ Cross section of probability wavefunctions of a nanoribbon with  infinitely long zigzag edges
along the x-axis in the presence of a perpendicular
magnetic field (a Landau gauge is used).
%Letters A  and B indicate whether A or B carbons are occupied.
 $|\psi_1|^2$ represents a localized surface state.  Two examples of LLL states  $|\psi_2|^2$ and $|\psi_3|^2$ are also shown.
These states all have nearly zero energies.} \label{wave}
\label{wave}
\end{center}
\end{figure}

We now explain qualitatively how mixed states of
Fig.\ref{fig:wavefunction1027zero}(c) and
Fig.\ref{fig:wavefunction1045zero}(c) can arise. An infinitely long
zigzag nanoribbon in a magnetic field has nearly zero energy surface
states that are localized on the edges in addition to ordinary
lowest Landau level states, see Fig.\ref{wave}.   The properties of
these states are given in Refs.\cite{Zh,Cas,Brey2,Brey3,Koh}: LLL
states of valley K (K$'$) are of B (A) type and localized surface
states  have a  mixed character between A and B. The surface states
can have various localization lengths but the minimum value is  of
order the carbon-carbon distance $a_0$\cite{Cas}. The armchair edges
couple K and K$'$ valleys\cite{Zh,Brey2,Cha}, and, consequently,
surface and LLL states of a nanoribbon can be coupled and give rise
to  mixed states with significant weight on the zigzag edges and
inside the dot, as shown in Fig.\ref{fig:wavefunction1027zero}(c)
and Fig.\ref{fig:wavefunction1045zero}(c). In addition, these mixed
states should display a significant occupation of both A and B
carbon atoms inside the dot since LLL states of different
chiralities are coupled by the armchair edges.  Our numerical result
is indeed consistent with this expectation, see
Fig.\ref{fig:wavefunction1027zero} (d).
%It also  suggests that these mixed states do not form readily in small dots and low magnetic fields.
As the ratio $\ell/L_{x,y}$ takes smaller values  the nature of these states become more like that of
LLL states (see  Fig.\ref{fig:edge2}).

We have investigated, in the presence of a magnetic field,   quasi-degenerate states of rectangular-shaped graphene
dots near the Dirac points. Some of these states are magnetic field independent surface states  while the other states
are field dependent.   We find numerically that the wavefunctions of these
states are all peaked near the zigzag edges with or without significant weight inside the dot.
The physical origin of the presence of a significant weight
 is the coupling between K and K' valleys due to the armchair edges.
This effect is expected to survive small deviations from perfect armchair edges as long as
they provide coupling between different valleys.
%In
%the limit of $L_x \rightarrow \infty$  we expect that the total
%electron density inside the dot will approach the bulk value of a nanoribbon with zigzag edges.
%Our scaling result, Fig.\ref{fig:scaling}, allows us to estimate  $N_D$ for a large dot at different magnetic fields, see
%Fig.\ref{fig:number_field}.
%Non-linear dependence on $B$ is visible.
%We note that the dependence of  $N_D$ on $B$ becomes more linear as the size of the
%dot and $B$ increases.
%Our scaling arguments  suggest that,   at magnetic fields of
%order $1T$ and
%dots of size of order  $100\times 100 nm^2$, it may be possible to measure a significant dependence of $N_D(\phi)$
%on $\phi$  ($\ell/L_{x,y}=0.26$).
%Our results show that  the initial non-linear dependence of  $N_D(\phi)$ on $B$ in the regime $2\ell\sim L_{x,y}$
%should turn into a linear dependence
%in the opposite regime  $\ell/L_{x,y}<<1$.
%(our estimate is
%$N_D\approx 8$ for $L_{x,y}=2000\AA$ and $\phi=0.0002 (B=20T)$.
%The number of  nearly zero energy states is  bigger for larger
%values $L_{x,y}$ and $B$.
Experimentally the  dependence of $N_D$ on $\phi$  may be studied by  measuring
STM properties\cite{Ys} or the optical absorption spectrum  as a function of magnetic
field\cite{Goer,Gusy,Hsu}.
In fabricating rectangular dots a special attention should be given to  the  direction of armchair edges since
the properties of dot may depend on it\cite{Ak}.

\begin{acknowledgments}
This work was supported by the Korea Research Foundation Grant funded by the Korean Government
(KRF-2009-0074470).    In addition this work was supported by the Second Brain Korea 21
Project.
\end{acknowledgments}


\begin{references}
\bibitem{Geim}
For recent reviews see: T. Ando, J. Phys. Soc. Jpn. {\bf 74}, 777 (2005); A. K. Geim and A. H.
MacDonald, Phys. Today {\bf 60} 35 (2007); A. H. Castro Neto, F. Guinea, N. M. R. Peres, K. S.
Novoselov, and A. K. Geim, Rev. Mod. Phys. {\bf 81}, 109 (2009).
\bibitem{Sci} K. S. Novoselov, A. K. Geim, S. V. Morozov, D. Jiang, Y.
Zhang, S. V. Dubonos, I. V. Grigorieva, and A. A. Firsov, Science
{\bf 306}, 666 (2004); H. Hiura, Appl. Surf. Sci. {\bf 222},
374(2004);
A. K. Geim and K. S. Novoselov, Nat.
Mater. {\bf 6}, 183 (2007).
\bibitem{Zhang} Y. Zhang, Y.W. Tan,  H.L. Stormer, and P. Kim,  Nature (London) {\bf 438}, 201 (2005).
\bibitem{Novo} K. S. Novoselov, A.K. Geim, S.V. Morozov, D. Jiang, M.I. Katsnelson, I.V. Grigorieva, S.V. Dubonos, and A.A. Firsov,
Nature (London) {\bf 438}, 197 (2005).
\bibitem{Mc} J.W. McClure, Phys. Rev.  {\bf 104}, 666 (1956).
\bibitem{Zh} Y. Zheng and T. Ando, Phys. Rev. B {\bf65}, 245420 (2002).
%\bibitem{She} L. Sheng, D. N. Sheng, F. D. M. Haldane, and Leon Balents,
%Phys.Rev.Lett. {\bf 99}, 196802 (2007).
%\bibitem{Chiu} Y. H. Chiu, Y. H. Lai, J. H. Ho, D. S. Chuu, and M. F. Lin, Phys. Rev. B {\bf 77}, 045407
%(2008).
\bibitem{Sh} S. G. Sharapov, V. P. Gusynin, and H. Beck, Phys. Rev. B {\bf 69}, 075104 (2004).
\bibitem{Fu} M. Fujita, K. Wakabayashi, K. Nakada, and K. Kusakabe, J. Phys. Soc. Jpn.  {\bf 65} ,1920 (1996).
\bibitem{Brey} L. Brey and H. A. Fertig, Phys. Rev. B {\bf 73}, 195408 (2006).
\bibitem{Son}Y. W. Son, M. L. Cohen, and S. G. Louie, Phys.Rev.Lett. {\bf
97}, 216803 (2006).
\bibitem{Brey2}  L. Brey and H. A. Fertig, Phys. Rev. B {\bf 73}, 235411 (2006).
\bibitem{Cas} A. H. Castro Neto, F. Guinea, and N. M. R. Peres, Phys. Rev. B {\bf 73}, 205408
(2006).
\bibitem{Gu} V. P. Gusynin, V. A. Miransky, S. G. Sharapov,
I. A. Shovkovy and C. M. Wyenberg, Phys. Rev. B {\bf 79}, 115431 (2009).
\bibitem{Ari} M. Arikawa, Y. Hatsugai, and H. Aoki, Phys. Rev. B {\bf 78},
205401 (2008).
\bibitem{Marti} A. DeMartino, L. Dell'Anna, and R. Egger, Phys. Rev. Lett. {\bf 98}, 066802 (2007).
\bibitem{Gia} G. Giavaras, P. A. Maksym, and M. Roy, J. Phys.: Condens. Matter {\bf 21}, 102201 (2009).
\bibitem{Haus} W. H\"{a}usler, and R. Egger, arXiv:0905.3667v1.
\bibitem{Sch} S. Schnez, K. Ensslin, M. Sigrist, and T. Ihn, Phys. Rev. B {\bf 78}, 195427 (2008).
\bibitem{ZZZ} Z. Z. Zhang, K. Chang, and F. M. Peeters, Phys. Rev. B {\bf 77}, 235411 (2008).
\bibitem{Rec} P. Recher, J. Nilsson, G. Burkard, and B.  Trauzettel, Phys. Rev. B {\bf 79}, 085407 (2009).
\bibitem{Gutt} J. G\"{u}ttinger, C. Stampfer, F. Libisch, T. Frey, J. Burgd\"{o}rfer, T.Ihn, and K. Ensslin, cond-mat arXiv: 0904.3506v2.
\bibitem{Fer} J. Fern$\acute{a}$ndez-Rossier and J. J. Palacios, Phys. Rev. Lett. {\bf
99}, 177204 (2007).
\bibitem{Ez} M. Ezawa,  Phys. Rev. B {\bf 76}, 245415 (2007).
\bibitem{Od} O. Hod, V. Barone, and G. E. Scuseria, Phys. Rev. B {\bf 77}, 035411 (2008).
\bibitem{Bj} B. Trauzettel, D. V. Bulaev, D. Loss
and G. Burkard, Nature Phys. {\bf 3}, 192 (2007).
\bibitem{Cha} C. Tang, W. Yan, Y. Zheng, G. Li, and
L. Li, Nanotechnology, {\bf 19}, 435401 (2008).
\bibitem{Ys} Y. Niimi, H. Kambara, T. Matsui, D. Yoshioka, and H.
Fukuyama, Phys. Rev. Lett. {\bf 97}, 236804 (2006).
\bibitem{Goer} M. O. Goerbig, J. -N. Fuchs, K. Kechedzhi, and V. I. Fal'ko, Phys. Rev. Lett. {\bf 99},
087402 (2007).
\bibitem{Gusy} V. P. Gusynin, S. G. Sharapov, and J. P. Carbotte, Phys. Rev. Lett. {\bf 98}, 157402
(2007).
\bibitem{Hsu} H. Hsu and L. E. Reichl, Phys. Rev. B {\bf 76}, 045418 (2007).
\bibitem{Koh}
M. Kohmoto and Y. Hasegawa, Phys. Rev. B {\bf76}, 205402 (2007).
\bibitem{Brey3} L.Brey and H.A. Fertig, Phys. Rev. B {\bf75}, 125434 (2007).
\bibitem{Ak}A. R. Akhmerov and C.W.J. Beenakker, Phys. Rev. B {\bf 77}, 085423 (2008).
\end{references}
\end{document}